\documentclass[twocolumn,superscriptaddress,showpacs,aps,floatfix]{revtex4-1}
\usepackage{amsmath,amsfonts,amssymb,epsfig,dcolumn,bm,dsfont,graphics,latexsym,color,graphicx}

\let\oldAA\AA

\renewcommand{\AA}{\text{\normalfont\oldAA}}
\newcommand{\ket}[1]{| {#1} \rangle} 
\newcommand{\aver}[1]{\langle {#1} \rangle} 
\begin{document}
\preprint{AIP/123-QED}
\title{Giant asymmetric proximity-induced spin-orbit coupling in twisted graphene/SnTe heterostructure}
\author{Marko Milivojevi\'c}
\email{milivojevic@rcub.bg.ac.rs}
\affiliation{Institute of Informatics, Slovak Academy of Sciences, 84507 Bratislava, Slovakia}
\affiliation{Institute for Theoretical Physics, University of Regensburg, 93053 Regensburg,
Germany}
\affiliation {Faculty of Physics, University of Belgrade, 11001 Belgrade, Serbia}
\author{Martin Gmitra}
\affiliation{Institute of Physics, Pavol Jozef \v{S}af\'{a}rik University in Ko\v{s}ice, 04001 Ko\v{s}ice, Slovakia}
\affiliation{Institute of Experimental Physics, Slovak Academy of Sciences, 04001 Ko\v{s}ice, Slovakia}
\author{Marcin Kurpas}
\affiliation{Institute of Physics, University of Silesia in Katowice, 41‑500 Chorz\'ow, Poland}
\author{Ivan \v Stich}
\affiliation{Institute of Informatics, Slovak Academy of Sciences, 84507 Bratislava, Slovakia}
\affiliation{Department of Natural Sciences, University of Saints Cyril and Methodius, 917 01 Trnava, Slovakia}
\author{Jaroslav Fabian}
\affiliation{Institute for Theoretical Physics, University of Regensburg, 93053 Regensburg, Germany}
\begin{abstract}
We analyze the spin-orbit coupling effects in a 3$^{\rm o}$-degree twisted bilayer heterostructure made of graphene and an in-plane ferroelectric SnTe, with the goal of transferring the spin-orbit coupling from SnTe to graphene, via the proximity effect. Our results indicate that the point-symmetry breaking due to the incompatible mutual symmetry of the twisted monolayers and a strong hybridization has a massive impact on the spin splitting in graphene close to the Dirac point, with the spin splitting values greater than 20 meV. 
The band structure and spin expectation values of graphene close to the Dirac point can be described using a symmetry-free model, triggering different types of interaction with respect to the threefold symmetric graphene/transition-metal dichalcogenide heterostructure.  
We show that the strong hybridization of the Dirac cone's right movers with the SnTe band gives rise to a large asymmetric spin splitting in the momentum space. Furthermore, we discover that the ferroelectricity-induced Rashba spin-orbit coupling in graphene is the dominant contribution to the overall Rashba field, with the effective in-plane electric field that is almost aligned with the (in-plane) ferroelectricity direction of the SnTe monolayer. 
We also predict an anisotropy of the in-plane spin relaxation rates. 
Our results demonstrate that the group-IV monochalcogenides MX (M=Sn, Ge; X=S, Se, Te) are a viable alternative to transition-metal dichalcogenides for inducing strong spin-orbit coupling in graphene.
\end{abstract}
\maketitle
\section{Introduction}
The field of spintronics~\cite{ZFS04,FME+07} has been providing fundamental science and technology with new functionalities derived from the electron spin as the fundamental degree of freedom. Two important interactions enable the efficient transfer of the information stored in the electron spin to electrically or magnetically readable one: spin-orbit and exchange coupling.

Its long spin-relaxation times~\cite{SKX+16,DFP+16} and high electronic mobility~\cite{BSJ+08,DSB+08} make graphene an excellent material for spin transport. However, to enable the manipulation of the spins in graphene and therefore truly unlock its full potential as a platform for spintronics, enhancing and controlling graphene’s spin-orbit coupling (SOC) and inducing magnetism is necessary. This can be achieved via the van der Waals proximity effects~\cite{GG13,ZMS+19,HMM+20,SFK+21,HZL+22,SBB+22,ZZN+22,ZRM+23}, which can successfully manipulate electronic, optical, magnetic, and spin properties of the target material.

It has already been demonstrated that graphene can be successfully proximitized so that it can become magnetic~\cite{HKG+14}. Similarly, strong SOC materials, such as transition metal dichalcogenides (TMDC)~\cite{ZHU,KZD+13,SYZ+13,KGR13,ABX+14,KBG+15}, can induce giant SOC into another two-dimensional material, without destroying its essential electronic properties. 
In graphene/TMDC heterostructures, giant spin lifetime anisotropy is observed~\cite{GAK+17,CGF+17,BST+18,ZCR+21}, defined as the ratio of lifetimes for spins pointing out of the graphene plane to those pointing in the plane. However, the preserved three-fold symmetry of graphene/TMDC heterostructures yields in-plane isotropy of spin relaxation rates. In contrast, heterostructures with no common symmetry are expected to induce fully three-dimensional spin relaxation anisotropy. This is a strong motivation to explore beyond the well-investigated graphene/TMDC stacks ~\cite{ATT+14,GF15,YLB+17,GF17,NZG+21,LSK+22,PDR+22,VPP+22,SMK+23} in the search for novel materials~\cite{IHS+19,TF23} that could be used as an alternative to induce both strong spin-orbit coupling and exotic spin textures in graphene. Particularly interesting are mixed-lattice heterostructures with incompatible symmetries, expected to give rise to novel spin textures. 

\begin{figure*}[t]
\centering
\includegraphics[width=0.88\textwidth]{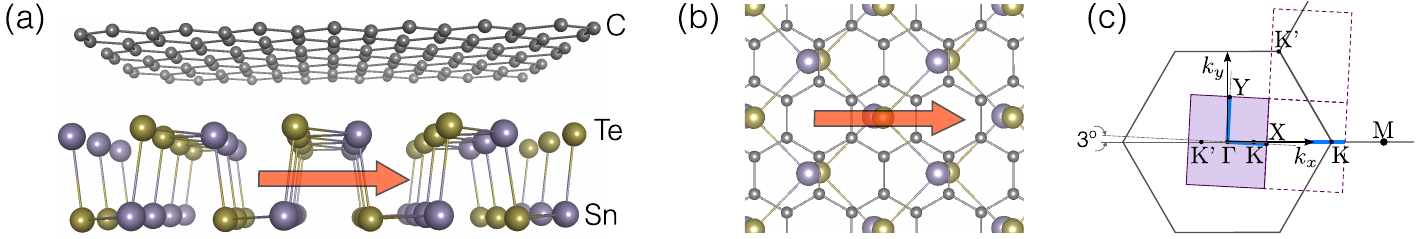}
\caption{Atomic structural model of the graphene/SnTe heterostructure: (a)~side view; (b)~top view, with the arrow indicating electrical polarization; (c) the first Brillouin zones of graphene and SnTe with high symmetry points and $k$-paths used for band structure calculations indicated by the blue lines.}\label{FigHet}
\end{figure*} 

In this paper, we investigate an example of a mixed-lattice heterostructure: hexagonal graphene and rectangular monolayer SnTe, which is a representative of group-IV monochalcogenides~\cite{GC15,FLL+15,GCN15,PVL19}. Monolayer SnTe has already been recognized as a material that can advance
spintronics applications~\cite{SCG+20,WGP+20}. 
It has strong SOC~\cite{AI19} and in-plane ferroelectricity~\cite{AMS+19,JOS+21} that is locked to the out-of-plane persistent spin texture.
Ferroelectricity in monolayer SnTe can be significantly modified by a substrate where the orbital hybridization and charge transfer could play vital roles \cite{Fu+19}.
We consider a small twist angle of three degrees between the bilayer constituents, giving rise to the symmetry-free environment of the proximitized graphene.
The performed DFT calculations and effective Hamiltonian modeling with no symmetry restrictions serve as proof that a Dirac cone nature of graphene is not destroyed due to the strong hybridization of graphene and SnTe bands, responsible for the observed giant asymmetric ($\propto 20\,$meV) spin splitting of graphene bands. Furthermore, analysis of the effective model reveals that the ferroelectricity-induced Rashba SOC is much stronger than the usual in-plane Rashba field, present due to the broken inversion symmetry. Finally, we predict that the in-plane spins of the graphene relax anisotropically due to the symmetry-free environment set up by the SnTe monolayer. 

This paper is organized as follows. In Section~\ref{Heterostructure} we describe the graphene/SnTe heterostructure, alongside the necessary details regarding the density-functional theory (DFT) calculation of the corresponding band structure. In Section~\ref{bandstructure}, we analyze the obtained band structure with and without SOC and describe the Hamiltonian model that can satisfactorily explain the DFT data. Furthermore, the impact of the obtained spin texture on spin relaxation is given in Section~\ref{spinrelaxation}, while the short conclusions are given in Section~\ref{Conclusions}.

\section{First-principles calculations details}\label{Heterostructure}
An atomic model of the graphene/SnTe heterostructure investigated in the paper is given in FIG.~\ref{FigHet}, wherein FIG.~\ref{FigHet}(a-b) the side and top views of the heterostructure are shown. In FIG.~\ref{FigHet}(c) the first Brillouin zones of graphene and SnTe are provided. The polarization direction coincides with the armchair direction of SnTe and is indicated by the orange arrow. The lattice parameter of graphene is equal to $a_0=2.46\AA$~\cite{HLS65}, while for the SnTe monolayer the values are $a=4.58\AA$ and $b=4.56\AA$, in line with the literature~\cite{AI19} (see Appendix~\ref{AppA} for more details). 

The commensurate heterostructure was obtained using the CellMatch code~\cite{L15} by straining the SnTe monolayer for $3.32$\%, containing 8 C, 2 Sn, and 2 Te atoms. The supercell lattice vectors ${\bm b}_1$ and ${\bm b}_2$ can be defined in terms of the graphene ones, ${\bm a}_1=a_0{\bm e}_x$ and ${\bm a}_2=a_0(-1/2{\bm e}_x+\sqrt{3}/2{\bm e}_y)$, as ${\bm b}_1=-3{\bm a}_1+10{\bm a}_2$ and  ${\bm b}_2=-{\bm a}_1+2{\bm a}_2$.
The relative twist angle of the armchair direction of the SnTe monolayer with respect to graphene is three degrees, appropriate for building a heterostructure with a small number of atoms. The nonzero twist angle between the bilayer constituents was chosen to break all the common symmetries between the graphene and SnTe monolayers and trigger the symmetry-free environment for graphene electrons. The applied strain could have a quantitative impact on the obtained results; nevertheless, our goal is to investigate to which level mixed-lattice heterostructures with no common symmetry could give rise to new physical phenomena.

We perform the electronic structure calculation of the twisted graphene/SnTe heterostructure by means of plane wave DFT suite Q{\sc{uantum}} ESPRESSO (QE)~\cite{QE1,QE2}, using the revised Perdew-Burke-Ernzerhof generalized gradient approximation for the exchange-correlation functional~\cite{PBEsol}. The kinetic energy cut-offs for the wave function and charge density were chosen to be 47\,Ry and 325\,Ry, respectively, assuming small Methfessel–Paxton energy level smearing~\cite{MP89} of 1 mRy. Additionally, $18\times 80$ $k$-points mesh for Brillouin zone integration was used, and a vacuum of 18\,${\AA}$ in the $z$-direction was assumed. The interaction between the monolayers was modeled using the
semiempirical Grimme-D2 van der Waals correction~\cite{G06,BCF+08}. The positions of atoms were relaxed using the quasi-Newton scheme using scalar-relativistic pseudopotentials, keeping the force and energy convergence thresholds for ionic minimization to $7\times10^{-4}$~Ry/bohr and $10^{-7}$ Ry, respectively. The average distance between the graphene plane and the closest SnTe planes (in $z$-direction) of the relaxed structure is equal to 3.35\,\AA.
For the self-consistent calculation including the spin-orbit coupling, the fully relativistic pseudopotential was used, keeping the same convergence threshold and $k$-mesh as in the case of the orbital relaxation. Also, dipole correction~\cite{B99} was applied to properly determine Dirac point energy offset due to dipole electric field effects.

We note that for convenience reasons the illustration of the band structure unfolded to the Brillouin zone of graphene is done with the DFT Vienna ab-initio simulation package VASP~6.2~\cite{KF96,KF99}, using as the input the relaxed structure from the QE code.

\section{Band structure analysis, effective model, and spin relaxation}\label{bandstructure}

\begin{figure}[t]
    \centering \includegraphics[width=0.38\textwidth]{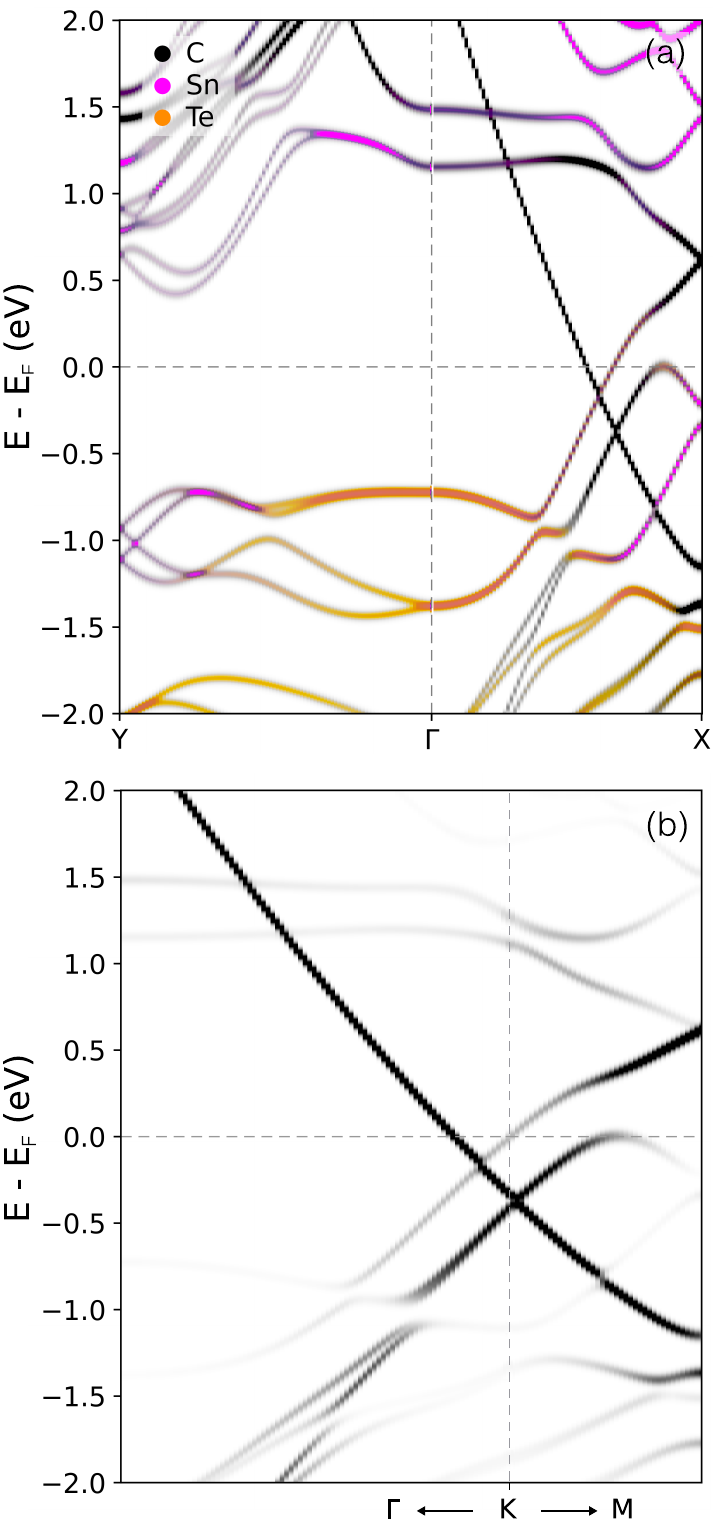}
    \caption{Calculated band structure for graphene/SnTe heterostructure along high symmetry points (a)~unfolded to Y$\Gamma$X path of SnTe Brillouin zone;
    (b)~unfolded to graphene Brillouin zone along the $\Gamma$KM path near the K.
    In~(a), 
       the spectral weights' color corresponds to the states' atomic projection: carbon is shown in black, tin in magenta, and tellurium in orange. In (b), only the spectral weight of carbon is given. 
       }
    \label{fig:bands}
\end{figure}

In FIG.~\ref{fig:bands} we present the band structure of the graphene/SnTe heterostructure unfolded to the Y$\Gamma$X path of the Brillouin zone of SnTe (FIG.~\ref{fig:bands}a) and $\Gamma$KM path of the Brillouin zone of graphene~(FIG.~\ref{fig:bands}b), with spectral weights of carbon presented in a) and b), and spectral weights of tin and tellurium presented in a). It is known that~\cite{AI19} SnTe is a semiconductor with two valence band maxima (pockets) located along the $\Gamma$X and $\Gamma$Y lines, close to the X and Y points, respectively. In FIG.~\ref{fig:bands}a), we notice a gap offset of SnTe along the $\Gamma$X line (electron doping) and strong hybridization with graphene bands.
Secondly, we note that the Dirac cone is electron-doped and is slightly shifted with respect to the K point. The K-point shift is an expected consequence of the broken three-fold rotational symmetry~\cite{SSL16}. Furthermore, the spectral weights of heterostructure atomic function reveal strong hybridization of the right movers of the Dirac cone with the ferroelectric SnTe bands, suggesting a very strong transfer of spin-orbit coupling from SnTe to graphene via the proximity effect. 
This assumption is consistent with DFT calculation, for which we derive an effective model and extract the relevant parameters that match the calculated band and spin structure.

\subsection{Effective model Hamiltonian}\label{ModelHamiltonian}
\begin{figure*}[htp]
\centering
\includegraphics[width=0.98\textwidth]{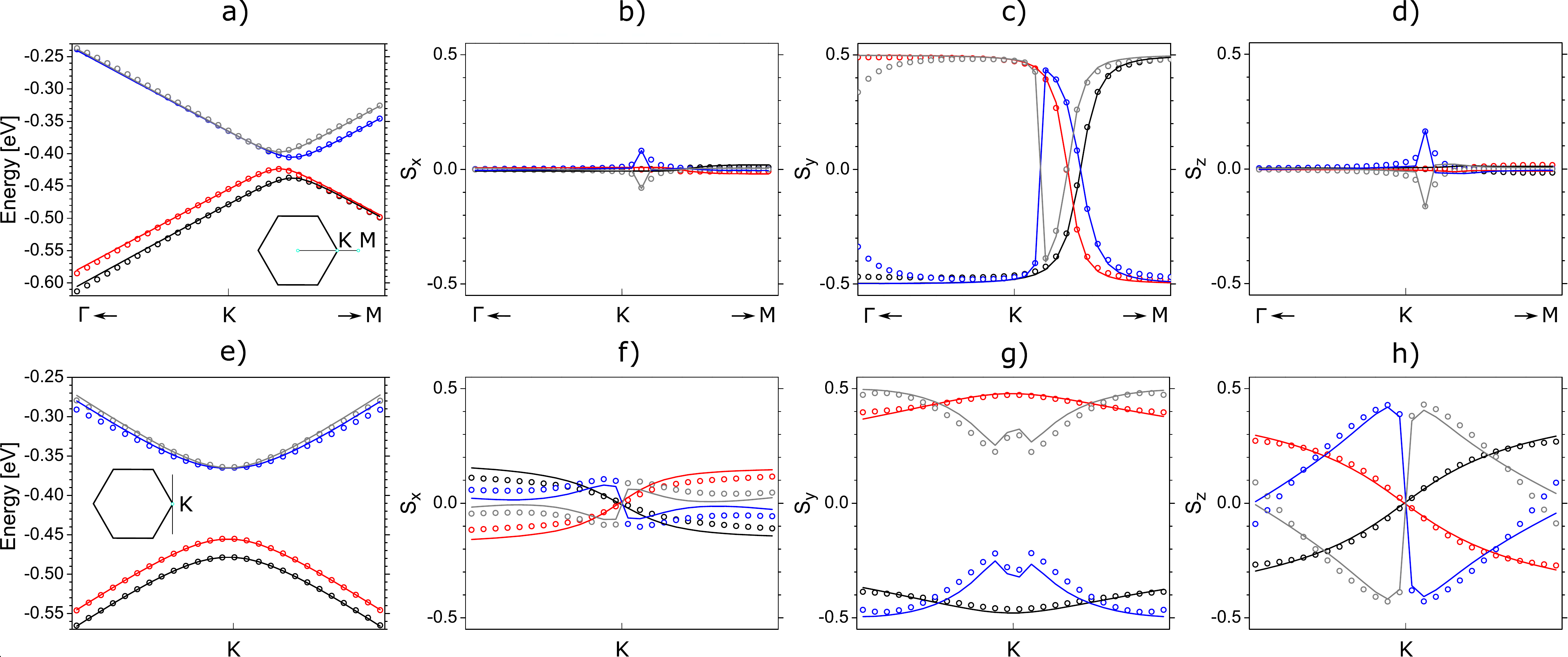}
\caption{Comparison between the DFT data and the fit obtained using the model Hamiltonian~\eqref{ModelHam} (parameters can be found in Table~\ref{FittingParameters}) in the vicinity of the $K$ point, $K=\frac{4\pi}{3a}(1,0)$, along different directions in the Brillouin zone:
in a) the band structure of the Dirac bands (labeled as C1, C2, V1, and V2) along the $\Gamma$KM path in the $x$-direction (see the inset) is given; the corresponding spin expectation values (in $\hbar$ units) are given in FIGS.~b)-d).
Furthermore, in FIGS.~e)-h) the k-path is chosen to be in the y-direction with the center at the K point.}\label{fit}
\end{figure*} 
To model the proximity-induced spin-orbit effects in graphene, we derive an effective Hamiltonian for the Dirac cone of graphene based on the symmetry of the studied heterostructure~\cite{KIF17,LK19,DRK+19}. As there is no restriction due to symmetry, we employ a general Hamiltonian form, spanning the graphene sublattice and spin spaces, restricted to terms linear in momenta (relative to K) 
${\bf k}$. 

Specifically, we model the Hamiltonian around the Dirac point as 
\begin{eqnarray}\label{ModelHam}
H&=&H_{\rm orb}+H_{\rm I}+H_{\rm R}+H_{\rm HI},
\end{eqnarray}
where $H_{\rm orb}$ represents the orbital Hamiltonian,
$H_{\rm I}$ is the intrinsic spin-orbit coupling Hamiltonian, and $H_{\rm R}$ is the Rashba spin-orbit coupling term always present when there is a structural asymmetry in the system. Finally, the hybridization-induced SOC term $H_{\rm HI}$ accounts for the SOC acquired due to the hybridization of the right movers of the Dirac cone with the SnTe bands. 

In more detail, the orbital Hamiltonian can be described as
\begin{eqnarray}\label{OrbitalHam}
H_{\rm orb}&=&\mu+\hbar v_{\rm F}(\tau k_x\sigma_x+k_y\sigma_y)+
\Delta\sigma_z+\Delta_{\rm s}\sigma_x,
\end{eqnarray}
where $\mu$ represents the chemical potential, $v_{\rm F}$ is the Fermi velocity, $\Delta$ represents the staggered potential, while $\Delta_{\rm s}\sigma_x$ is the term responsible for the shift of the Dirac cone in the $x-$direction due to the broken out-of-plane rotational symmetry of the graphene/SnTe heterostructure, giving rise to the uniaxial deformation of the graphene lattice~\cite{SSL16}. Finally, $\tau=1\, (-1)$ corresponds to the K (K') point and $k_x$ and $k_y$ are the Cartesian components of the electron wave vector measured from K (K').

Next, we assume the presence of the sublattice-dependent intrinsic SOC which can be written as~\cite{GF15}
\begin{eqnarray}\label{intrinsicSOC}
H_{\rm I}&=&\lambda_{\rm I}^{A}[(\sigma_z+\sigma_0)/2]\tau S_z+\lambda_{\rm I}^{B}[(\sigma_z-\sigma_0)/2]\tau S_z,
\end{eqnarray}
where $\lambda_{\rm I}^{A/B}$ represent the strength of the intrinsic SOC parameters on the sublattice $A/B$, while $S_z$ is the Pauli z-matrix acting in the spin-subspace.

Next, we discuss the effects of momentum-dependent Rashba spin-orbit coupling. 
In the general case, the gradient of the crystal potential $\nabla V ({\bf r})$
can have all three components. The z-component gives rise to the usual Rashba spin-orbit field
$\propto k_xS_y-k_yS_x$, while x- and y- components of the crystal potential gradient can generate the novel terms $k_yS_z$ and $k_xS_z$, respectively, which can be interpreted as the $x-$ and $y-$ component of the effective electric field. The appearance of the novel Rashba spin-orbit terms is to be expected since SnTe is a strong in-plane ferroelectric material (in the armchair~\cite{AMS+19,JOS+21} direction of the SnTe monolayer), suggesting a strong gradient of the crystal potential in the polarization direction. The three-degree twist angle between the polarization (armchair) direction of the SnTe monolayer and the $x$-axis of the heterostructure's coordinate frame indicates that both in-plane components of the crystal potential gradient should be present.
Thus, we can model $H_{\rm R}$ as
\begin{eqnarray}\label{Rashba}
H_{\rm R}&=&\sum_{i=0}^3 \tau^{\delta_{i,2}}\sigma_i\Big[\beta_1^{i}k_xS_z+\beta_2^{i}k_yS_z+\lambda^{i}(k_xS_y-k_yS_x)\Big]\nonumber\\
\end{eqnarray}
additionally assuming the general pseudospin dependence of the Rashba SOC Hamiltonian, justified due to the absence of common symmetry between the constituents in the studied heterostructure. Finally, let us mention that interaction $\propto k_x S_z$ and $k_y S_z$ (in-plane Rashba field terms) give rise to the out-of-plane spin texture. On the other hand, the terms $\propto k_x S_y-k_y S_x$ are called the out-of-plane Rashba field terms,  giving rise to the in-plane spin texture. In~\eqref{Rashba},  $\delta_{i,j}$ represents the Kronecker delta function, being equal to 1 when $i=j$, and 0 otherwise. Thus, $\tau^{\delta_{i,2}}$ is equal to $\tau$ for $i=2$ and 1 otherwise.

Finally, to model the momentum-independent hybridization of the right movers of the Dirac cone with the SnTe bands, we assume the presence of the momentumless hybridization-induced  SOC term 
\begin{eqnarray}
H_{\rm HI}&=& \sum_{i=0}^3 \tau^{1+\delta_{i,2}}\sigma_i(\alpha^{i}_1S_x+\alpha^{i}_2 S_y),
\end{eqnarray}
where the parameters $\alpha^{i}_{1/2}$, $i=0,...3,$ represent the pseudospin-dependent components of the spin-orbit field, while $\delta_{i,j}$ is the Kronecker delta function, as in~\eqref{Rashba}.

\subsection{Model parameters}\label{Results}
In order to extract the model parameters, we have performed DFT calculations with QE focusing on the Dirac cone structure of graphene bands. Comparison between the DFT data and the fit obtained using the model Hamiltonian~\eqref{ModelHam} can be found in FIG.~\ref{fit}.
More precise, in FIG.~\ref{fit} a)-d) band structure and spin expectation values of the graphene bands around the Dirac cone along the $\Gamma$KM path in the $x$-direction is given ($k_y=0$), using the values of $k_x$ in the range $k_x\in(-\kappa,\kappa)$, measured from the K point, where $\kappa=0.026\,\AA^{-1}$. In FIG.~\ref{fit} e)-h), the band structure and spin expectation values are given by assuming the k-path in the y-direction of the graphene Brillouin zone ($k_x=0$), where $k_y\in(-\kappa,\kappa)$ is the distance with respect to the K point (see inset).

The obtained parameters are gathered in Table~\ref{FittingParameters}. The Fermi velocity, in this case, is equal to $0.76\times10^6$ m/s, which is of the same order of magnitude as in other proximitized graphene heterostructures~\cite{YSM+12,JZ21,TJU+22}.
Other spin-independent terms consist of the chemical potential $\mu$, staggered field potential $\Delta\sigma_z$, followed by the Dirac cone direction shift in the x-direction, $\Delta_{\rm s}\sigma_x$. 

When discussing the sublattice-resolved spin conserving SOC parameters $\lambda_{\rm I}^{A}$ and $\lambda_{\rm I}^{B}$, it is to be mentioned that it can be written as $\lambda_{\rm I}^{A/B}=\lambda_{I}\pm\delta\lambda_{\rm I}$, where 
$\lambda_{I}=14.46$\,meV is called the intrinsic or Kane-Mele SOC term~\cite{CY62,KM05}, while $\delta\lambda_{\rm I}=-0.05$\,meV is the valley-Zeeman term~\cite{GF15,SMK+23}. Whereas the intrinsic SOC Hamiltonian gives rise to the out-of-plane spin texture, the in-plane hybridization-induced field induces the in-plane spin texture. Within the pseudo-spin dependent terms, we can identify the dominant ones, proportional to $(\sigma_0+\sigma_1)S_y$, since $\alpha_2^0=\alpha_2^2$. A simple numerical analysis of the model Hamiltonian shows that the term $\alpha_2^0(\sigma_0+\sigma_1)S_y$ affects only the right movers of the Dirac cone and orients the spin in the y-direction, with the size of the spin splitting proportional to the parameter $\alpha_2^0$. The appearance and the dominant influence of such a term can be traced to the strong hybridization of the right movers of the Dirac cone with SnTe bands, crucially affecting the shape of the Dirac bands. Moreover, since the SnTe band is parallel to the right movers of the Dirac cone over a wide range of ${\bf k}$, it was natural to assume that the k-independent term was to describe such an interaction. 

Finally, we discuss the Rashba SOC. As mentioned in  Sec.~\ref{ModelHamiltonian}, the gradient of the crystal potential in the $x/y/z$ direction gives rise to the Rashba terms $\propto k_yS_z/k_xS_z/(k_xS_y-k_yS_x)$, whose influence is quantified by the pseudospin-dependent parameters $\beta_2^i/\beta_1^i/\lambda^i$, $i=0,..,3$. To get a deeper insight into the nature of the effective electric field generating the Rashba interaction, we define a generalized pseudospin-independent Rashba coupling strength terms as $|\beta_{1/2}|=\sqrt{\sum_{i=0}^3 (\beta_{1/2}^{i})^2}$ and $|\lambda|=\sqrt{\sum_{i=0}^3 (\lambda^{i})^2}$. 
Using this formula and the fitted parameters, we obtain $\beta_1=47.61$ meV\,\AA, $\beta_2=368.73$ meV\,\AA, $\lambda=155.60$ meV\,\AA. 
We see that the Rashba coupling parameters are significant, but still much smaller than the Rashba parameter in SnTe (1.23 eV $\AA$~\cite{AI19}).  Furthermore, the main contribution to the overall Rashba spin-orbit coupling comes from the in-plane components, suggesting the ferroelectric origin of the Rashba SOC. 
By comparing the ratio of $\beta_1$ and $\beta_2$, the 7.35 angle of the in-plane electric field generating Rashba is determined, being almost aligned (3 degrees) with the polarization direction of the SnTe monolayer. This confirms the assumption of the ferroelectric nature of the in-plane Rashba SOC field.

\begin{table}[t]
\caption{Fitting parameters of the graphene bands in the vicinity of the K point. Chemical potential is given as $\mu$, 
Fermi velocity as $v_{\rm F}$, staggered potential as $\Delta$, while the strain-induced K point shift is equal to $\Delta_{\rm s}$. Sublattice-dependent intrinsic spin-orbit coupling is represented as
$\lambda_{\rm I}^{A/B}$; also, we label the pseudo-spin dependent Rashba parameters as $\beta_{1/2}$ and $\lambda$, with components $\beta_{1/2}^{i}$ and $\lambda^{i}$  ($i=0,...,3$) spanning the pseudo-spin space. Finally, $\alpha_{1/2}$ represent the in-plane HI field, with pseudo-spin components $\alpha_{1/2}^i$, $i=0,...,3$.}\label{FittingParameters}
\centering
\begin{tabular}{cc}\hline\hline
$v_{\rm f}$[$10^6$m/s]&0.76\\
$(\mu,\Delta,\Delta_{\rm s})$\,[meV]& $(-416.03,0.18,-48.54)$ \\ 
$(\lambda_{\rm I}^A,\lambda_{\rm I}^B)$\,[meV]& $(14.41,14.51)$ \\ 
$\beta_1$\,[meV\,\AA]& $(0.97,1.51,-27.30,-38.97)$ \\ 
$\beta_2$\,[meV\,\AA]& $(-116.50,-349.00,1.94,-24.17)$ \\
$\lambda$\,[meV\,\AA]&$(-42.77,-5.85,-4.48,149.42)$ \\
$\alpha_{1}$\,[meV]& $(0.05,0.08,-1.52,-2.17)$ \\ 
$\alpha_{2}$\,[meV]& $(5.67,5.67,0.11,-1.35)$
\\\hline\hline
\end{tabular}
\end{table}
\begin{figure}[t]
\centering
\includegraphics[width=0.49\textwidth]{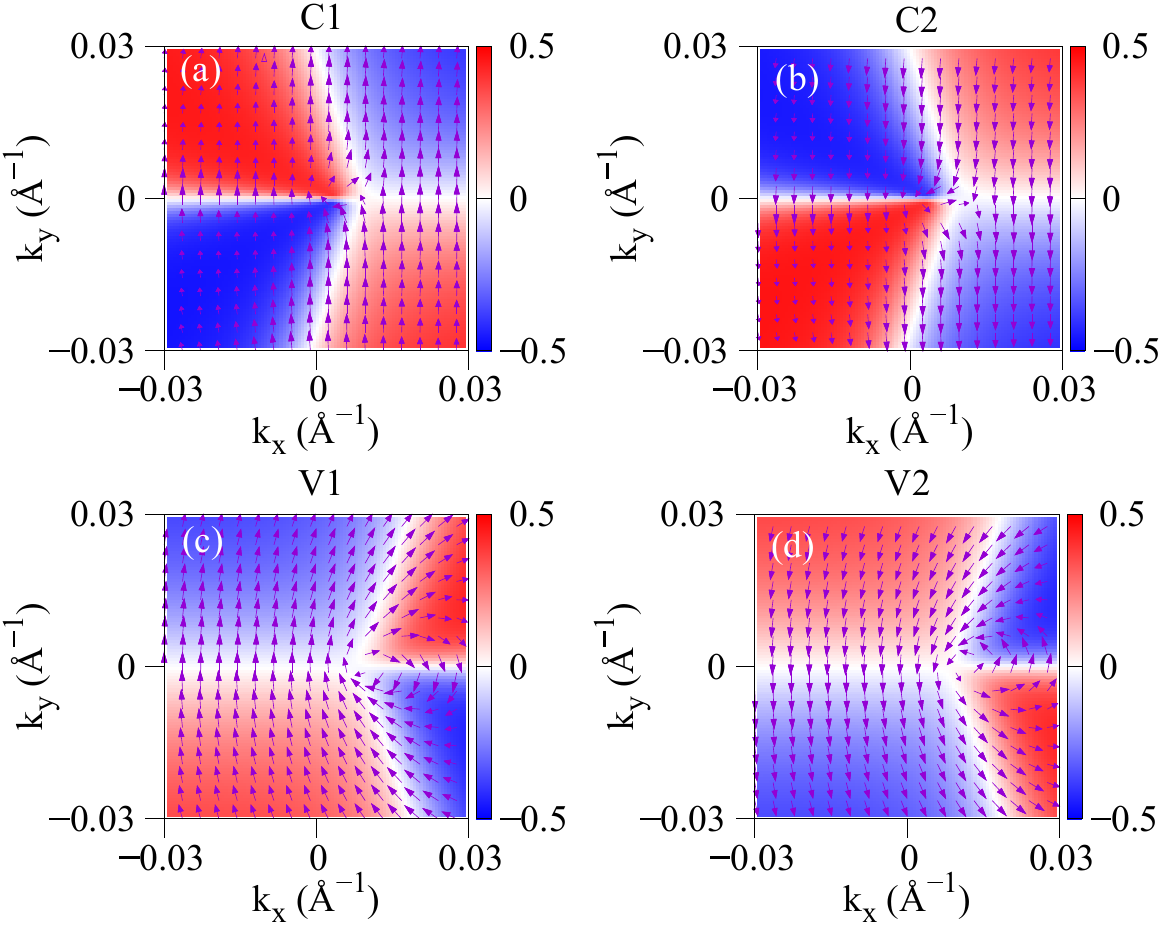}
\caption{Spin texture of the spin split Dirac cone bands around the $K=4\pi/3a(1,0)$ point is given for each of the four bands. The values of ${k_x/k_y}\in(-0.03,0.03)\AA^{-1}$ are measured with respect to the K point. The arrows represent the in-plane spin texture, while the out-of-plane spin texture is represented by the color scale.}\label{SpinTexture}
\end{figure} 

The spin texture of the proximitized graphene of Dirac bands (C1, C2, V1, and V2, see FIG.~\ref{fit}a) is given in  FIG.~\ref{SpinTexture}, assuming that the $k_x/k_y\in(-0.03,0.03)\AA^{-1}$ is the distance measured from the $K=4\pi/3a(1,0)$ point. First, one notices that the in-plane texture, represented in FIG.~\ref{SpinTexture} by arrows, is almost fully polarized in the y-direction, representing a signature of the hybridization-induced SOC term $\propto(\sigma_0+\sigma_1)S_y$, responsible for the highly asymmetric band structure and the giant (bigger than 20\,meV) spin splitting of the Dirac bands. On the other hand, the z component of spin is influenced by both the intrinsic SOC $H_{\rm I}$ and the in-plane Rashba SOC Hamiltonian terms.
This was concluded indirectly by comparing the averaged z component of the spin-orbit field (see below); however, a similar conclusion stems from the comparison of the DFT data for $S_z$ with the Hamiltonian model~\eqref{ModelHam} that uses the parameters from Table~\ref{FittingParameters}
while manually turning off the in-plane Rashba coupling parameters
$\beta_{1/2}^i$ in one case and intrinsic SOC parameters $\lambda_{\rm I}^{A/B}$ in the second case.

\subsection{Proximity-induced spin relaxation}\label{spinrelaxation}
Finally, we show that the novel spin texture induced into graphene by the interaction with the SnTe monolayer has a genuine fingerprint on the spin relaxation rate anisotropy, not observed before. In the case of twisted graphene/TMDC heterostructure, the common ${\bf C}_3$ symmetry of the heterostructure gives rise to the in-plane isotropy of spin relaxation rates and the giant in-plane/out-of-plane spin relaxation rate anisotropy~\cite{CGF+17,GVC+18,ZGF19}. In the case of the graphene/SnTe heterostructure, the absence of common symmetry between the heterostructure constituents triggers new SOC terms and as a consequence, new channels for spin relaxation.
Assuming the Dyakonov-Perel~\cite{DP72} mechanism of spin relaxation, which should be dominant in proximitized graphene~\cite{ZCR+21}, spin relaxation rates can be described as
\begin{eqnarray}\label{relaxtimes}
 \tau_x^{-1}&=&\tau_p(\aver{\Omega^{2}_{{\bf k},y}}_{\rm p}+\aver{\Omega^{2}_{{\bf k},z}}_{\rm p})+
\tau_{\rm iv}(\aver{\Omega^{2}_{y}}_{\rm iv}+\aver{\Omega^{2}_{z}}_{\rm iv}),\nonumber\\ 
 \tau_y^{-1}&=&\tau_{\rm p}(\aver{\Omega^{2}_{{\bf k},x}}_{\rm p}+\aver{\Omega^{2}_{{\bf k},z}}_p)+
\tau_{\rm iv}(\aver{\Omega^{2}_{x}}_{\rm iv}+\aver{\Omega^{2}_{z}}_{\rm iv}),\nonumber\\ 
 \tau_z^{-1}&=&\tau_{\rm p} (\aver{\Omega^{2}_{{\bf k},x}}_{\rm p}+\aver{\Omega^{2}_{{\bf k},y}}_{\rm p})+\tau_{\rm iv}(\aver{\Omega^{2}_{x}}_{\rm iv}+\aver{\Omega^{2}_{y}}_{\rm iv}),\nonumber\\
\end{eqnarray}
where $\tau_{i}$, $i=x,y,z$, represents a lifetime of spins pointing in the $i$ direction; $\tau_{\rm p}$ is the momentum
relaxation time, $\tau_{\rm iv}$ is the intervalley scattering time, while
$\aver{\Omega_{i}^2}_{\rm iv}$ and $\aver{\Omega_{{\bf k},i}^2}_{\rm p}$ are the Fermi contour averages of the momentum-independent (index iv) or momentum-dependent (index p) effective spin-orbit field components. By definition, the average $\aver{\Omega^2}$ of an arbitrary spin-orbit field component $\Omega^2$ is defined as
\begin{equation}\label{averagedOmega}
    \aver{\Omega^2}=\frac{1}{\rho(E_F)S_{BZ}}\int_{FC}\frac{\Omega^2}{\hbar |v_F(k)|}dk,
\end{equation}
where $\rho(E_F)$ represents the density of states per spin at
the Fermi level, $v_F(k)$ is the Fermi velocity, while $S_{BZ}$ is the area of the Fermi surface. 

To calculate the momentum-dependent and momentum-independent spin-orbit fields, one has to turn on/off the corresponding interactions of the graphene Hamiltonian~\eqref{ModelHam}. Detailed, in the case of the z-components of the spin-orbit field, terms proportional to $\lambda_{I}^{A/B}$ are momentum-independent, while the terms proportional to $\beta_{1/2}^{i}$, $i=0,...,3$, are momentum-dependent; for the in-plane spin-orbit field components, one has to distinguish between contributions of the momentum-independent terms proportional to $\alpha_{1/2}^{i}$, $i=0,...,3$, and the momentum-independent terms proportional to $\lambda^{i}$, $i=0,...,3$.
The procedure to calculate the spin-orbit fields for valence and conductance band can be found in Appendix~\ref{AppB}. Assuming the Fermi level $E_{\rm F}$ in a range from -170\,meV to -50\,meV below the valence band maximum, the spin-orbit field averages $\aver{\Omega^{2}_{x/y/z}}_{\rm iv}$ and 
$\aver{\Omega^{2}_{{\bf k},x/y/z}}_{\rm p}$ are calculated for the valence bands. The results that can be found in Appendix~\ref{AppB} show that the dominant contribution to the spin relaxation rate comes 
from $\aver{\Omega^{2}_{y}}_{\rm iv}$. When compared to $\aver{\Omega^{2}_y}_{\rm iv}$, the z-components
$\aver{\Omega^{2}_{{\bf k},z}}_{\rm p}$ and $\aver{\Omega^{2}_{z}}_{\rm iv}$
are also sizable for bigger $|E_{\rm F}|$, while for smaller $|E_{\rm F}|$,  $\aver{\Omega^{2}_{x}}_{\rm iv}$ should be taken into the account. 
To estimate the impact of the averaged spin-orbit field components on the spin relaxation rates, the relative ratio  $\tau_{\rm iv}/\tau_{\rm p}$ has to be known.
Although this value is hard to determine from the experiments, it can be estimated to be within the range $\tau_{\rm iv}/\tau_{\rm p}\in(10-60)$~\cite{ZCR+21}.

\begin{figure}[t]
    \centering    \includegraphics[width=0.75\columnwidth]{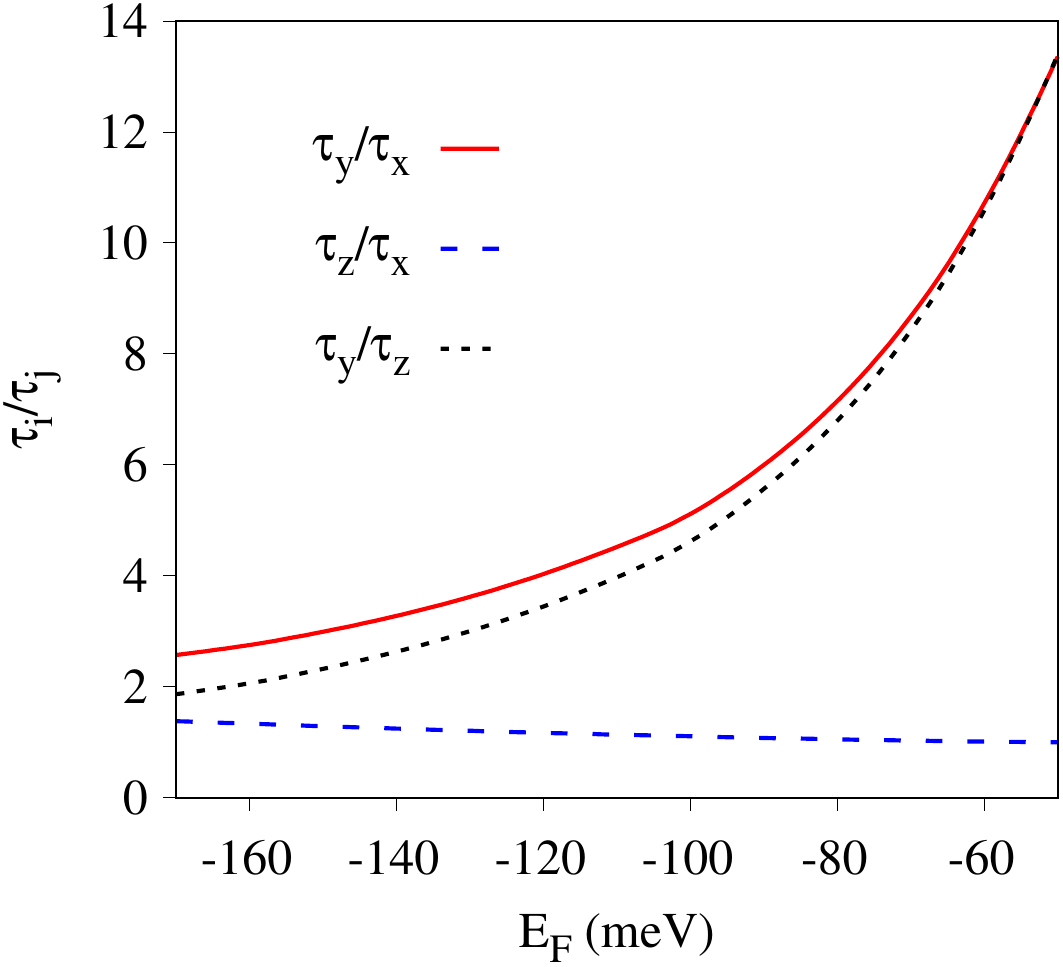}
    \caption{Dependence of anisotropies of Dyakonov-Perel spin relaxation times $\tau_y/\tau_x$,  $\tau_z/\tau_x$, and $\tau_y/\tau_z$ on the position of the Fermi level $E_F$. In the calculations, the relation between the
    momentum relaxation time $\tau_{\rm p}$ and the intervalley scattering time $\tau_{\rm iv}$ is set to be
    $\tau_{\rm iv}=10\tau_{\rm p}$. 
    }
    \label{spinrel}
\end{figure}

In FIG.~\ref{spinrel}, we analyze the ratio of spin relaxation rates $\tau_y/\tau_x$, $\tau_z/\tau_x$, and $\tau_y/\tau_z$ on the doping level, being in the range from -170\,meV to -50\,meV below the valence band maximum, assuming  $\tau_{\rm iv}=10\tau_{\rm p}$.
Our calculations have shown that the results are independent of the ratio $\tau_{\rm iv}/\tau_{\rm p}\in(10-60)$. This is a clear consequence of the two facts, the dominant roles of $\tau_{\rm iv}$ time scale and averaged momentum-independent spin-orbit fields appearing in~\eqref{relaxtimes}. Focusing on the spin relaxation rates, the results demonstrate the prominent in-plane spin relaxation anisotropy, while the ratio $\tau_z/\tau_x$ is proportional to 1, suggesting that the dominant time scale is given by $\tau_y$. This is dramatically different from the calculated~\cite{CGF+17} and observed~\cite{GAK+17} spin relaxation rates in graphene/TMDC heterostructures, 
where $\tau_x=\tau_y\ll \tau_z$, suggesting the importance of substrate in inducing and manipulating different spin properties in graphene.

\section{Conclusions}\label{Conclusions}
We analyzed the proximity-induced spin-orbit effect in a heterostructure made of graphene and SnTe monolayer, assuming a relative twist angle of three degrees between the monolayers.
Focusing on the band structure of graphene close to the Dirac cone,
we show that, due to the strong hybridization with SnTe bands,  giant asymmetric spin splitting of the graphene bands occurs. We show that the data from the ab-initio calculations can be fitted to the symmetry-free model and discuss the novel terms present in the effective Hamiltonian due to the broken symmetry.
We show that the hybridization-induced spin-orbit coupling induces strong spin-splitting asymmetry, having its origin in the hybridization of the right movers of the Dirac cone with SnTe bands. Moreover, the strong in-plane ferroelectricity of the SnTe monolayer is transferred to the graphene bands via the Rashba effect, whose in-plane field is much stronger than the more common out-of-plane field, also present in graphene/TMDC heterostructures. Finally, we report large in-plane anisotropy of spin relaxation rates in graphene bands, present due to the broken symmetry of the studied heterostructure. Thus, we show that group-IV monochalcogenide MX (M=Sn, Ge; X=S, Se, Te) monolayers can be an efficient alternative to transition-metal dichalcogenides for inducing strong spin-orbit coupling into a desired material and novel patterns of spin relaxation rates.

\acknowledgments
M.M. acknowledges the financial support
provided by the Ministry of Education, Science, and Technological Development of the
Republic of Serbia and DAAD Research Grant 57552336. This project has received
funding from the European Union's Horizon 2020 Research and Innovation Programme under the Programme SASPRO 2 COFUND Marie Sklodowska-Curie grant agreement No. 945478.
M.G.~acknowledges financial support provided by the Slovak Research and Development Agency provided under Contract No. APVV-SK-CZ-RD-21-0114 and by the Ministry of Education, Science, Research and Sport of the Slovak Republic provided under Grant No. VEGA 1/0695/23 and Slovak Academy of Sciences project IMPULZ IM-2021-42 and project FLAG ERA JTC 2021 2DSOTECH.
M.K.~acknowledges financial support provided by the National Center for Research and Development (NCBR) under the V4-Japan project BGapEng V4-JAPAN/2/46/BGapEng/2022.
I.{\v S}~acknowledges financial support by APVV-21-0272, VEGA 2/0070/21, VEGA 2/0131/23, and by H2020 TREX GA No. 952165 project.
J.F. acknowledges support from
EU project 101135853 (2DSPIN-TECH) and FLAG ERA
JTC 2021 2DSOTECH.
The authors gratefully acknowledge the Gauss Centre for Supercomputing e.V. for funding this project by providing computing time on the GCS Supercomputer SuperMUC-NG at Leibniz Supercomputing Centre.

\appendix
\section{Crystal parameters of SnTe monolayer}\label{AppA}

The crystal parameter determination was obtained by minimizing the total energy of the system in the absence of the spin-orbit coupling within the plane wave QE package, using the SG15 Optimized Norm-Conserving Vanderbilt (ONCV) pseudopotentials~\cite{H13,SG15}, with the kinetic energy cut-offs for the wave function and charge density 80\,Ry and 320\,Ry, respectively. The force and energy convergence thresholds were set to $10^{-5}$~Ry/bohr and $10^{-8}$ Ry, respectively.  For the Brillouin zone integration, $10\times 10\times 1$ $k$-points mesh was considered using the Monkhorst-Pack scheme; a vacuum of 20\,${\AA}$ in the $z$-direction was used. The obtained crystal parameters are $a=4.58\AA$ and $b=4.56\AA$, in line with~\cite{AI19}.

Since an important characteristic of a SnTe monolayer is its ferroelectricity~\cite{CLL+16,SCG+20,WGP+20}, we also calculate this property of a system using the determined crystal parameters. In this case, the fully relativistic SG15 ONCV pseudopotentials with the same energy cut-offs for the wave function and charge density as in the non-relativistic case were used, with the $k$ point mesh $25\times 25\times 1$. By performing the Berry phase calculation~\cite{KV93}, the polarization value of 23.39 $\mu C/{\rm cm}^2$ in the armchair direction of the SnTe monolayer was obtained when the effective thickness 1~nm for the monolayer is used, consistent with~\cite{WLX+14}.

\section{Derivation of the spin-orbit field in proximitized graphene}\label{AppB}
\begin{figure}
    \centering
    \includegraphics[width=0.98\columnwidth]{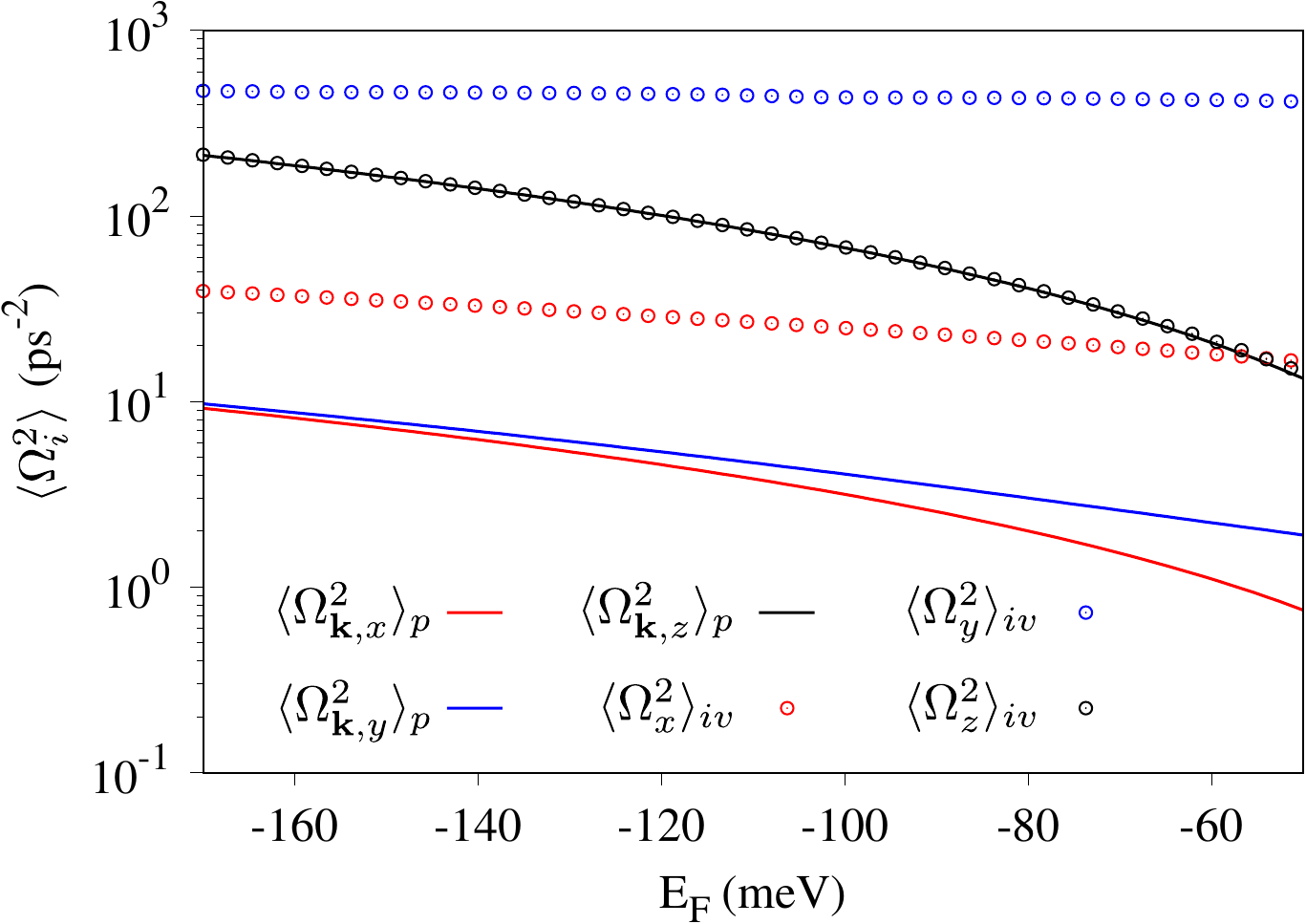}
    \caption{Fermi contour average of the momentum-dependent $\aver{\Omega^{2}_{{\bf k},x/y/z}}_{\rm p}$ and momentum-independent $\aver{\Omega^{2}_{x/y/z}}_{\rm iv}$ spin-orbit fields versus the position of the Fermi level calculated.}
    \label{fig:fs_omega2}
\end{figure}
Here we show how the spin-orbit field ${\bm \Omega}_{\bf k}$ in proximitized graphene~\cite{ZCR+21} can be derived straightforwardly from the effective Hamiltonian H~\eqref{ModelHam}. 
As a first step, the orbital part of the Hamiltonian, $H_{\rm orb}$~\eqref{OrbitalHam}, is diagonalized,
\begin{equation}
    H_{\rm orb}\Psi=E\Psi, 
\end{equation}
where $\Psi=\{\ket{\Psi_{CB1}},\ket{\Psi_{CB2}},\ket{\Psi_{VB1}},\ket{\Psi_{VB2}}\}$ describe eigenvectors of conductance and the valence bands of graphene. As the next step, the effective Hamiltonian~\eqref{ModelHam} is written in the basis $\Psi$ and has the following form
\begin{equation}
    \mathcal{H}=\begin{pmatrix}
     \mathcal{H}_{CC}     & \mathcal{H}_{CV}  \\
      \mathcal{H}_{VC}   &  \mathcal{H}_{VV}
    \end{pmatrix},
\end{equation}
where $ \mathcal{H}_{CV/VC}$ represents the interaction between the conductance and the valence bands. Finally, the Hamiltonian~$ \mathcal{H}$ is projected onto the conductance or valence band only, taking into account the interaction between them, 
\begin{eqnarray}
    \mathcal{H}_{CB}&\equiv&\Big[\mathcal{H}_{CC}+ \mathcal{H}_{CV}  (E-  \mathcal{H}_{VV})^{-1}\mathcal{H}_{VC}\Big],\nonumber\\
     \mathcal{H}_{VB}&\equiv&\Big[\mathcal{H}_{VV}+ \mathcal{H}_{VC}  (E-  \mathcal{H}_{CC})^{-1}\mathcal{H}_{CV}\Big].
\end{eqnarray}
The obtained conductance/valence band Hamiltonians $\mathcal{H}_{CB/VB}$ can be written in the following form
\begin{equation}
    \mathcal{H}_{CB/VB}=\epsilon_0 \mathcal{I}_2+\frac{\hbar}2\begin{pmatrix}
        \Omega_Z   & \Omega_X-\rm{i}\Omega_Y \\
     \Omega_X+\rm{i}\Omega_Y  &  -\Omega_Z
    \end{pmatrix},
\end{equation}
where $\mathcal{I}$ is the identity 2$\times$2 matrix, $\epsilon_0$ is the orbital energy, while ${\bm \Omega}=(\Omega_X,\Omega_Y,\Omega_Z)$ corresponds to the spin-orbit field, representing the system dependent quantity that crucially affects the spin relaxation times of valence and conductance bands of graphene.  

Using the third-order perturbation theory we have taken into account the interaction between the conductance and valence bands and calculated the Fermi contour averages of the momentum-independent and momentum-dependent spin-orbit fields $\aver{\Omega^{2}_{x/y/z}}_{\rm iv}$ and 
$\aver{\Omega^{2}_{{\bf k},x/y/z}}_{\rm p}$, as in~\eqref{averagedOmega}.
In our calculations, we have assumed the Fermi level $E_{\rm F}$ values from -170\,meV to -50\,meV below the valence band maximum. This justifies the usage of the perturbation approach since there is a sizable gap between the valence and conductance bands. In FIG.~\ref{fig:fs_omega2}, the dependence of the Fermi contour averaged spin-orbit field on $E_{\rm F}$ is given,  showing that the dominant spin relaxation channel is  $\aver{\Omega^{2}_y}_{\rm iv}$, while  $\aver{\Omega^{2}_{{\bf k},z}}_{\rm p}$ and $\aver{\Omega^{2}_{z}}_{\rm iv}$
are comparable for bigger $|E_{\rm F}|$; for smaller $|E_{\rm F}|$  $\aver{\Omega^{2}_{x}}_{\rm iv}$ should also be taken into the account.

\end{document}